\documentclass[aps,apl,twocolumn,showpacs,superscriptaddress]{revtex4}
\usepackage{graphicx}
\usepackage{graphics}
\usepackage{epsf}
\begin{document}

\title{Correlation of electron-phonon in cubic zirconia: $G_0W_0 @DFPT$ scheme }
ý\author{Zohreh Dabiri}ý

ý\affiliation{Physics Department, Yazd University, P.O. Box 89195-741, Yazd, Iran.}
ý\author{Ali Kazempour}ý

ý\affiliation{Department of Physics, Payame Noor University ,P.O. BOX 119395-3697, Tehran, Iran}
ý\affiliation{Nano Structured Coatings Institute of Yazd Payame Noor university, P.O. Code 89431-74559, Yazd, Iran.}
ý\author{Mohammad Ali Sadeghzadeh}ý

ý\affiliation{Physics Department, Yazd University, P.O. Box 89195-741, Yazd, Iran.}
\begin{abstract}

The indirect carrier-phonon vertex function is evaluated for cubic zirconia by employing the many-body perturbation theory on top of the frozen phonon model. The phonon-induced carrier velocity and spectral renormalization are represented at each individual lattice vibronic mode. Our results show that for the indirect electron-phonon scattering, optical phonons make up the dominant contribution. It is observed that, despite the phonon-phonon interaction, there is a substantial polaronic character for all the activated lattice optical vibrations which is in accordance with the previous experimental and theoretical finding concerning the crucial role of optical phonons. Eventually our results are capable to discuss that, in the presence of optical phonons a transition between small to large polaron is probable.
\end{abstract}

\maketitle

\section{INTRODUCTION}

In semiconductors, insulators, and semi-metals with a small electron density of states at the Fermi level, lattice vibronic states (phonon) are the dominant quasi-particles that are responsible for the electronic and thermal conductivity. Accordingly, studying the phonon-induced effects in materials has attracted immense attention. These interactions are identified to determine the highest operation limits of any electronic device. Strikingly, high dielectric materials are new emerging candidates for improving and modifying the electronic and thermal conductivity in the new generation of electronic devices\cite{kralik1997,p36,p37}. Among a wide variety of these materials, zirconia($ZrO_{2}$) is of significant scientific and technological importance due to its superior mechanical stability and outstanding electronic and thermal properties\cite{kissi,coating,china}. One of the encouraging applications of $ZrO_{2}$ is its potential ability to be replaced by $SiO_{2}$, which is a conventional gate oxide of field effect transistors. $ZrO_{2}$ has a high dielectric constant $(\varepsilon\sim 25)$ with a large band gap(5-6eV). Compared with the conventional silicon-based gate oxides, $ZrO_{2}$ prevents leakage current more effectively\cite{leakage}.
There are many theoretical and experimental studies in electronic structure and lattice vibrations of different phases of $ZrO_{2}$\cite{kralik1997,parlinski,19973,car,goff,stampfl,namavar}. However, in contrast to the electron-electron(e-e) coupling, there are only a few available theoretical and experimental studies on the electron-phonon(e-ph) interaction in $ZrO_{2}$. For instance, an \emph{ab initio} investigation of the phonon dispersion and the lattice contribution to the dielectric properties of zirconia is reported in Ref.\cite{zhao}. The calculated Born effective charge in their study indicates the crucial role of the phonon vibrations to the static dielectric constant $\epsilon_{0}$, which is higher for the cubic phase in comparison to the other zirconia phases.
In Ref.\cite{sternik,grun}, using density functional perturbation theory(DFPT), a Longitudinal-Transverse optical(LO-TO) mode splitting at $\Gamma$ point which is qualitatively large, is reported. This effect is a result of the non-analytic term that requires the knowledge of the Born effective charge tensors and dielectric constant of $ZrO_{2}$. According to their study, when taking phonon-phonon (ph-ph) coupling into account, the phonon softening is inevitable and the reported Gruneisen Parameter also reflects the requirement of anharmonic approximation for the precise calculation of electronic and thermal transport. In this regard, the variation of phonon distribution should change the electronic distribution, effectively. All the aforementioned arguments declare relevant clues about strong e-ph interaction in zirconia.
Recently, the electronic correlation of $ZrO_{2}$ in the framework of many-body perturbation theory in GW approximation\cite{p33,fermiliquid,stefano} is explored in Ref.\cite{jiang}. Their results address the interplay between the volume deformation potential(hydrostatic strain) and the electronic degree of freedom. They discussed the fact that, the dielectric permittivity tensor of zirconia is dominated by the static contribution($\epsilon_{0}$) while, the electronic part ($\epsilon_{\infty}$) is a factor 5-6 times smaller than $\epsilon_{0}$. Such a big discrepancy results in the emergence of a considerable electric polarization, particularly due to the LO phonons. This phenomenon highlights the e-ph coupling, which is neglected in the study of Jiang \emph{et al.}. In Ref.\cite{goodphonon}, the obtained results from an \emph{ab initio} calculations for dielectric function showed a good accordance with the experimental results, since they considered the electron-optical phonon contribution to evaluate the dielectric function of $ZrO_{2}$.
Despite all difficulties, there are numerous experiments on the electronic structure and phase transition properties of $ZrO_{2}$ compounds. The importance of the phonon scattering and its influence on the thermal conductivity of the $ZrO_{2}$ microstructure is discussed in Ref.\cite{good0}, where the change in the surface strain was measured. In another experimental study using ellipsometry, the obtained dielectric function of polycrystalline Yttria stabilized zirconia(YSZ) thin films illustrated consequential electron coupling with the optical phonons\cite{Schmidt}. Nevertheless, a comprehensive study that focuses on e-ph scattering effects on  $ZrO_{2}$ electronic features, is still lacking\cite{jiang} and the present work may compensate it for future studies in this field.

In this study, we measure the vibronic states of cubic $ZrO_{2}$ through indirect phonon activation in the framework of the frozen phonon model.
Our work has the advantage of being compared and scaled with those applications in which the e-ph interaction plays a consequential role in tuning the efficiency of electronic devices through managing quantities such as the figure of merit. This paper is structured as follows: In Sec. II we will explain the theoretical approach, i.e. a brief theory of all quantities which are discussed in this study, and details of our technical computing. In Sec. III we will reveal and discuss our findings with the inclusion of the combined many-body and DFPT, henceforth denoted $G_0W_@DFPT$ and emphasize the electronic and thermal features of zirconia focusing on its cubic lattice phase. Section IV summarizes the main points and findings.

\section{\textbf{COMPUTATIONAL METHOD}}

Within the state-of-the-art density functional perturbation theory approach\cite{stefano}, the direct e-ph vertex functions using the Migdal approximation\cite{migdal} is:
\begin{equation}
\Lambda_{n\acute{}n}^{\lambda}(k,q)=\sqrt{\frac{\hbar}{2\omega_{\lambda,q}}}\sum_{ij}\varepsilon_{i,j}^{\lambda}(q)
\frac{1}{\sqrt{M_{i}}}\langle\psi_{n\acute{},k+q}|\frac{\partial U_{scf}}{\partial u_{qij}}|\psi_{n,k}\rangle
\label{eq2naft}
\end{equation}
where $\psi_{n,k}$ and $\psi_{n\acute{},k+q}$ are the electronic Bloch states, $U_{scf}$ is the self-consistent Kohn-Sham potential created by all electrons and ions, $\varepsilon_{i,j}^{\lambda}(q)$ are the eigen vector components of the dynamical matrix correspond to the displacement of i-th ion(mass $M_{i}$) in the direction j due to the phonon mode $(\lambda,q)$, and $u_{q,i,j}\equiv N^{-1}\sum_{p}exp(iq.R_{l})u_{lij}$, where $u_{lij}$ denotes the displacement of i-th ion from unit cell l in the direction j and $R_{l}(l=1,2,...,N)$ denotes the unit cell.
Since the direct calculations of the e-ph coupling matrix using the aforementioned method requires lots of time consuming computations, we apply many-body perturbation theory in the GW approximation\cite{p32,p33,hedin,Hybertsen,p49,p50} by employing the displacement snapshots which are derived from DFPT in the framework of the frozen phonon picture. Thus $G_0W_0 @DFPT$ is capable of measuring the strength of the indirect e-ph coupling and carrier scattering in cubic $ZrO_{2}$. Having earned the e-e correlation in each individual snapshot lead us to the qualitative prediction of the coupling between the electronic cloud and phonon degrees of freedom. Phonon branches comprise nine modes arranged in three longitudinal and six transverse parts for acoustic and optical phonons. Treating each phonon as a perturbation of the self-consistent Kohn-Sham potential is crucial to our theoretical study.

\begin{figure}
\begin{center}
\includegraphics[angle=0,width=0.45\textwidth]{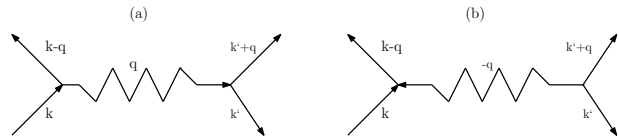}
\end{center}
\caption{\label{electron}The phonon induced electron-electron interaction( phonon emission(a) and phonon absorption(b)).
}
\end{figure}

Fig.\ref{electron} represents the indirect e-ph interaction schematically, one electron of wave vector (k) emits a phonon of wave vector (q) and scatters into the state of $(k-q)$, then a phonon of wave vector (q) immediately absorbs by another electron of wave vector $(k\acute{})$ that scatters into the state $(k\acute{}+q)$\cite{grosso}. Refer to Fig.\ref{electron}, the associated electronic states and their related energies are:
\begin{eqnarray}
|i\rangle &=& |k,k';0\rangle\nonumber\\
|\alpha\rangle &=& |k-q,k';1\rangle\nonumber\\
|f\rangle &=& |k-q,k'+q;0\rangle\nonumber\\
E_{i} &=& E_{k}+E_{k'}\nonumber\\
E_{\alpha} &=& E_{k-q}+E_{k'}+\hbar\omega_{q}\nonumber\\
E_{f} &=& E_{k-q}+E_{k'+q}\nonumber
\end{eqnarray}
assuming $\hbar\omega_{q}\approx\hbar\omega_{D}$ where $\omega_{D}$ is Debye frequency, it is deduced that within the energy shell of the order of $\hbar\omega_{D}$ around Fermi energy, this process leads to an attractive interaction between electrons. The indirect e-ph coupling strength is given by:
\begin{eqnarray}
\langle f | H_{indirect}| i \rangle &=& |M_{q}|^{2}[\frac{\hbar \omega_{q}}{(E_{k}-E_{k-q})^{2}-\hbar^{2} \omega^{2}_{q}}+ \nonumber\\
&&\frac{\hbar \omega_{q}}{(E_{k'+q}-E_{k'})^{2}-\hbar^{2} \omega^{2}_{q}}]
\end{eqnarray}
where, the scattering amplitude for electron transition from the state (k) into the state (k-q) with the creation of a phonon (q) on the vacuum phonon state is $M_{q} \langle 1|a_{q}\dag  | 0 \rangle=M_{q}$, similarly the scattering amplitude from the state $(k')$ into the state $(k'+q)$ with the absorption of a phonon (q) is: $ M_{q}^{*} \langle 0 | a_{q} |1 \rangle =M_{q}^{*}$

In many-body calculations, to investigate the electronic properties of the systems, the quasi-particle self-energy has the leading contribution. In this scheme, the carrier lifetimes or equivalently damping rates can be extracted from the imaginary part of the self-energy\cite{hedin,lifetime,GW2lt,r6lt}:
\begin{equation}
\tau^{-1}=-2\int dr \int dr' \psi_{i}(r) Im \Sigma(r,r';\varepsilon_i)\psi_{i}(r')
\label{ta}
\end{equation}
where $\psi_{i}(r)$ and $\varepsilon_i$ are the electronic states and the related energies, respectively. According to Hedin\cite{hedin} along with the GW on-shell approximation\cite{lifetime} the self-energy expands at the first order in the dynamically screened interaction:
\begin{equation}
\Sigma(r,r':\varepsilon_i)=\frac{i}{2\pi}\int d\varepsilon_i' e^{-i\delta^+\varepsilon_i'} G(r,r':\varepsilon'_i-\varepsilon_i)W(r,r':\varepsilon_i')
\label{sigma}
\end{equation}
in which, G is the Green function of the electronic system, W is the dynamically screened Coulomb interaction and $\delta^+$ denotes a positive
infinitesimal time. In this approximation, by neglecting the off-diagonal terms in the self-energy equation one can determine the poles of the Green function that result in the lifetimes of the electronic excitations:
\begin{eqnarray}
\Delta\Sigma_{ii}(\omega)=\int dr\int dr'\psi_i^*(r)[\Sigma(r,r',\omega)-v_{LDA}(r)\nonumber\\
\delta(r-r')]\psi_i(r')
\end{eqnarray}
in this equation $v_{LDA}$ denotes the local one electron potential, and the electron damping rate follows as:
\begin{equation}
\tau^{-1}=-2Z_i Im\Delta \Sigma_{ii}(\varepsilon_i),
\label{damp}
\end{equation}

The quasi-particle spectral residue or energy renormalization factor(Z-factor) is an important quantity to evaluate the e-ph coupling strength in the interacting many-body systems and is defined as $Z_{n}(\textbf{k})\equiv |\langle \Psi_{nk}|\psi_{nk}\rangle|^{2}$, where $|\psi_{nk}\rangle$ is the bare electron state in the \emph{n}-th Bloch band and $|\Psi_{nk}\rangle$ is that of the perturbed system. The Z-factor characterizes the e-ph interaction strength and assesses how much quasi-particle energies are close to the DFT values. Moreover, in the case of no correlations the Z-factor is equal to the unity and generally defined as:
\begin{equation}
Z_i=\lbrace 1-\frac{\partial Re \Delta \Sigma_{ii}(\omega)}{\partial\omega}|_{\omega=\varepsilon_i}\rbrace
\label{z}
\end{equation}
The quantity $-\frac{\partial Re \Delta \Sigma_{ii}(\omega)}{\partial\omega}|_{\omega=\varepsilon_i}$ is a parameter which denotes the strength of the e-ph coupling. Thus a lower Z-factor indicates a stronger e-ph coupling.

Another important feature of our many-body calculations is the velocity renormalization factor which corresponds with the real part of the self-energy and is related to the e-ph coupling strength through:
\begin{equation}
1-Z_{i}^{-1}=\frac{v_{i}(k)-v_{0}(k)}{v_{i}(k)}
\label{vel}
\end{equation}
in this equation $v_{i}(k)$ and $v_{0}(k)$ denote the electron velocity in the non-interacting system and in the presence of e-ph interaction, respectively.

In following we will emphasize our computational details to achieve the e-ph coupling which causes the renormalization effects and eventually results in polaronic features in our system. To perform the frozen phonon calculations in the $G_0W_0 @DFPT$ framework, we used the QUANTUM ESPRESSO package\cite{espresso}, and to improve the quasi-particle properties, the dielectric matrix was calculated using the general plasmon-pole approximation as implemented in the Yambo package\cite{yambo}. Valence electronic wavefunctions are expanded in a plane-wave basis that is truncated at a cutoff energy of 50 Ry. The electronic and phonon calculations were performed on a uniform $6\times6\times6$ momentum grid, and the ultra-soft Martins-Troullier norm-conserving pseudo-potential with Perdew-Wang LDA exchange and correlation were applied for both Zirconium and Oxygen atoms.

\section{RESULTS AND DISCUSSIONS}

\subsection{frozen phonon calculations}

\begin{figure}
\begin{center}
\includegraphics[angle=0,width=0.48\textwidth]{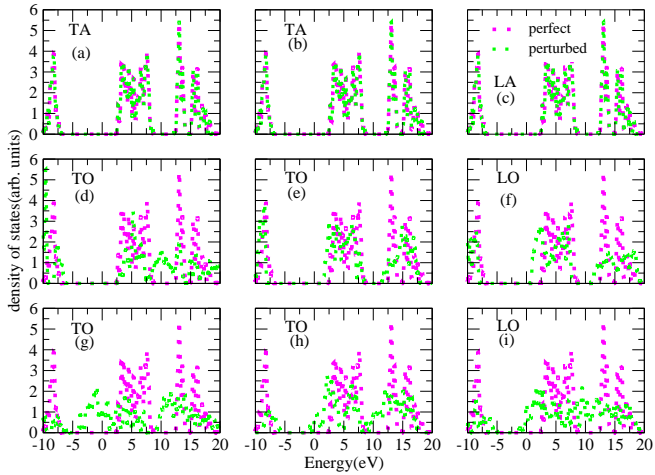}
\end{center}
\caption{\label{moldelectrondos} (color online)Cubic zirconia density of states versus electron energy. Energy redistribution occurs due to the phonon induced e-e interaction.
}
   \end{figure}

Fig.\ref{moldelectrondos} shows the $G_0W_0@DFPT$ results for zirconia electronic density of states in which the unperturbed DOS is located in the background of each panel. The acoustic branches (a), (b), and (c), declare no deviation from the unperturbed DOS. In the other words, due to the presence of acoustic phonons, ions displaced in-phase. Hence, acoustic phonons have no effect on the energy redistribution of the electrons. To put it another way, the interactions between acoustic modes and electrons are qualitatively energy conservative and the correlation between the electrons as well as the coherency of the system, being maintained. As a consequence, to randomize the electron velocity due to the elastic scattering, is the only effect of electron-acoustic phonon coupling\cite{cardonabook}.
Panels (d), (e), (f), (g), (h), and (i) of Fig.\ref{moldelectrondos} depict the density of states for the six perturbed lattice vibrations in the presence of optical phonons. The electron-optical phonon scattering is an inelastic interaction; therefore, the electrons relax by emitting a LO or TO phonon as is illustrated in Fig.\ref{electron}. These inelastic scatterings cause considerable redistribution of the carrier momenta and energies as is shown in Fig.\ref{moldelectrondos}. Zirconia is an ionic material; therefore, the Frohlich interaction is expected to play the main role in the scattering processes. This interaction is proportional to the $[\omega_{LO}(\epsilon_{\infty}^{-1}-\epsilon_{0}^{-1})]^{1/2}$\cite{cardonabook}. As it it stated before, the ratio of $\epsilon_{0}$ and $\epsilon_{\infty}$  varies between 5 and 6 orders of magnitude, that confirms the relativity of the scattering processes to the presence of LO phonons. Apparently in optical phonon branches, the density of state satellite character is destroyed drastically from the perfect background, except for the panels (e) and (f). In these two modes, the band-gap of $ZrO_{2}$ is still clear, but smaller than the perfect DOS; therefore, $ZrO_{2}$ behaves like a narrow band-gap semiconductor.

\begin{figure}
\begin{center}
\includegraphics[angle=0,width=0.48\textwidth]{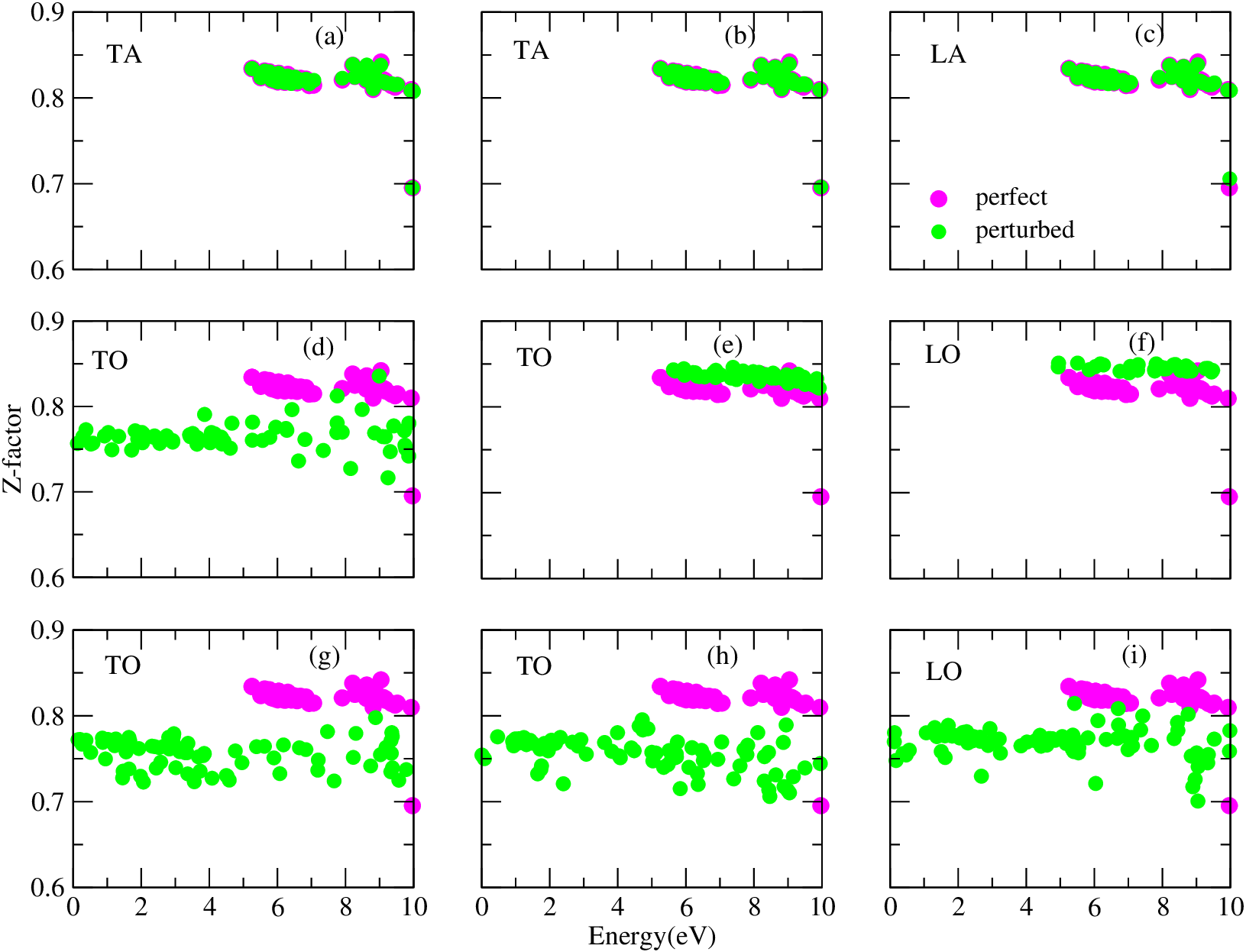}
\end{center}
\caption{\label{renormalization} (color online) Renormalization factor due to the phonon induced e-e coupling. Renormalization in optical branches is sufficient for carriers to have polaronic character except for configurations (e) and (f).
}
\end{figure}

Fig.\ref{renormalization} shows the Z-factor for the conduction states in cubic $ZrO_2$. In the acoustic modes no discrepancy is observed compared to the background as is observed in the DOS pattern; therefore, no appreciable momentum redistribution occurs during the acoustic phonon induced e-e interaction. Two results are depicted from the optical modes: $1)$ the second TO and the first LO modes in panels (e) and (f) exhibit an increase in energy renormalization, contrary to the other optical modes, and $2)$ these two branches maintain the symmetry of the crystal field. The Z-factor is a measure of correlation between an individual electron with the rest of the electronic cloud. As it is argued in Ref.\cite{Hybertsen}, the typical Z-factor values of the transition metal oxides vary from 0.5 to 0.8 for d and s orbitals and it is well known that a higher Z-factor results in a lower correlation between electrons. From this point of view, compared to the other optical modes, it can be elucidated from snapshots (e) and (f) that, the e-ph coupling is reduced, indicating a higher values for electronic mobility. The calculated Z-factor for (e) and (f) show that, the phonon-induced electron renormalization is insufficient for these electrons to have polaronic character. In other optical modes, based on the comparison between binding energy and the zirconia band-width together, with a reduction in the Z-factor, result in the polaronic nature of electronic transport in cubic $ZrO_2$. The considerable reduction in the renormalization weight (Z-factor), confirms the obtained results in the Ref.\cite{goodphonon} in which the dielectric matrix of $ZrO_2$ is reproduced by taking the phonon self-energy along with the electron self-energy, into account and also improvement in the obtained screening length by taking e-ph coupling into account, is in good agreement with the available experimental values\cite{goodphonon}.
Inferred from our calculations, it is also predictable that, increasing the Z-factor value in panels (e) and (f), and then decreasing drastically in (g), (h), and (i), results in a small to large polaron transition by the substantial role of the optical phonons as is clarified in Ref.\cite{polaron}. The second effect of those phonons which break the crystal symmetry and remove the degeneracy of cubic $ZrO_2$, is that electrons in our system to behave like electron gas with lacking the single quasi-particle character and simultaneously following the Fermi liquid model of quadratic pattern for the scattering rates Ref.\cite{fermiliquid}.

\begin{figure}
\begin{center}
\includegraphics[angle=0,width=0.48\textwidth]{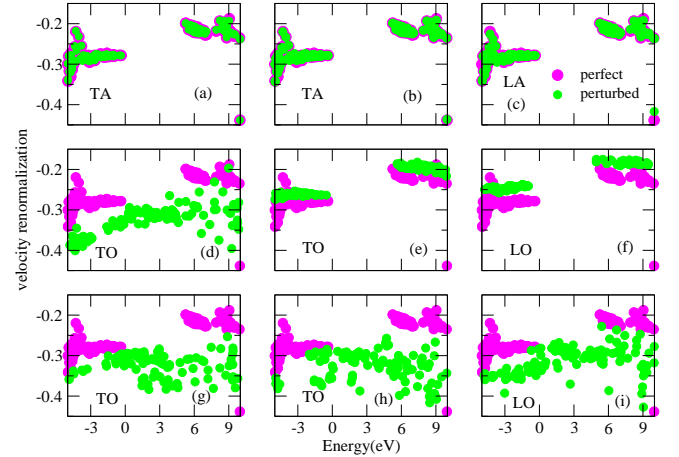}
\end{center}
\caption{\label{velocity} (color online)Velocity renormalization versus electron energy. Polaronic and non-polaronic features for carrier transport is deduced from the velocity renormalization curves as is described in the text.
}
\end{figure}

The calculated velocity renormalization of electrons is reported in Fig.\ref{velocity}. In acoustic branches we observe no important changes in the velocity renormalization factor which means that the activation of the acoustic modes during in-phase ionic displacements should not affect the momentum redistributions of electrons. Furthermore, the out-phase ionic displacements in (e) and (f), result in the velocity renormalization promotion,  confirming a non-polaronic nature in the first LO and second TO snapshots, consistent with the DOS and Z-factor results. There is a considerable reduction in velocity renormalization in the last three optical modes which indicates the polaronic feature of the carrier transports.

\begin{figure}
\begin{center}
\includegraphics[angle=0,width=0.45\textwidth]{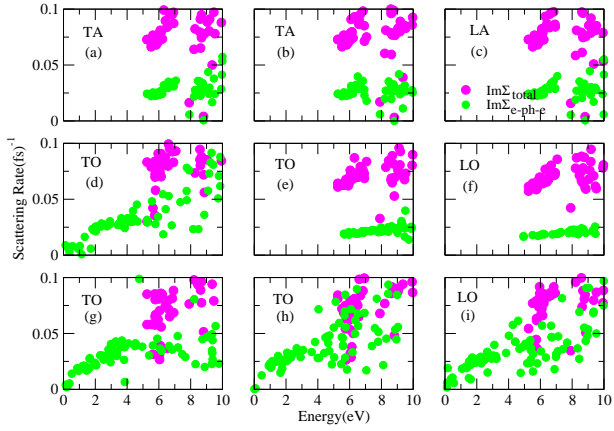}
\end{center}
\caption{\label{matt} (color online) Calculated scattering rates for total contribution(e-e and e-ph-e) and the e-ph-e contribution versus energy.
}
  \end{figure}

Fig.\ref{matt} shows the obtained total scattering rates of $(Im\sum_{e-e}+Im\sum_{e-ph-e})$ according to the Matthiessen's rule together with individual e-e intermediated a phonon scattering rate $(Im\sum_{e-ph-e})$, where $Im\sum_{e-e}$ and $Im\sum_{e-ph-e}$ denote scattering rates for $ZrO_2$ at equilibrium structure and phonon-displaced snapshots, respectively. From the branches (d), (g), (h), and (i) of Fig.\ref{matt} compared to  the e-ph-e scattering rates in each vibronic modes, it is deduced that the contributions of the e-ph coupling play the main role in the total scattering rate in which cubic $ZrO_{2}$ behaves metal-like having free carriers whereas, in panels (e) and (f) a weak contribution to the total scattering rates is illustrated, indicating a weak e-ph-e coupling in these two modes as we observed from the foregoing results. Moreover, there are small gaps and fluctuations in all configurations that can be attributed to the shape of the electronic band-structures. It is observed that in the higher conduction bands, there is a small band spread from point W to L while the band-width in the rest of the Brillouin zone, especially from $\Gamma$ to X, experiences a wider energy variation, which is the cause of the decay in fluctuation rates. It is worth mentioning that our frozen phonon calculations are performed at the $\Gamma$ point of the Brillouin zone where cubic $ZrO_{2}$ experiences its stable phase and does not show any soft mode\cite{parlinski,sternik}.

\begin{figure}
\begin{center}
\includegraphics[angle=0,width=0.48\textwidth]{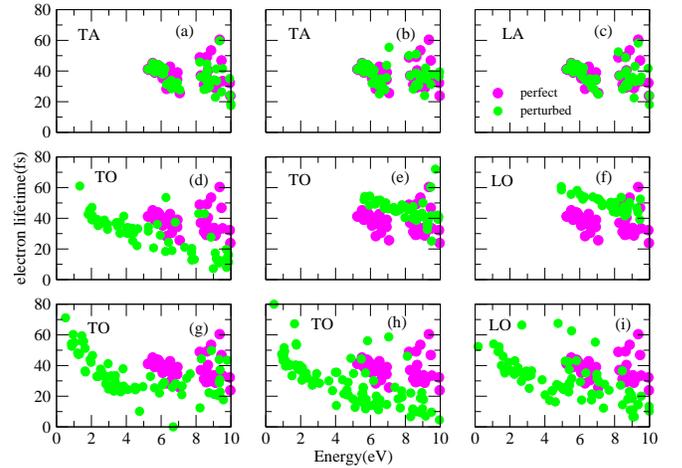}
\end{center}
\caption{\label{lifetime2} (color online) Electron lifetimes versus energy. Contrary to the optical modes, the calculated coupling between electron-acoustic phonons has no important effects on the electron lifetimes. The observed lifetime pattern in optical modes follow Fermi liquid model except for the (e) and (f) modes.
}
  \end{figure}

Fig.\ref{lifetime2} shows the obtained lifetimes for each phonon mode compared to the equilibrium state of $ZrO_{2}$. In acoustic branches, despite the electron assisted acoustic phonon correlations, electron lifetimes remain unaltered, which is in line with our foregoing results. In fact, in these modes electron lifetimes for inter-band and intra-band transitions have the same order of magnitude, though inter-band transitions due to the optical phonons are slightly larger than intra-band transitions due to the acoustic phonons.
In optical branches, two effects are distinguished: $1)$ lifetimes in panels (d), (g), (h), and (i) decrease and follow the quadratic pattern of Liquid Fermi model and $2)$ lifetimes in panels (e) and (f) increase. From the latter effect together our previous findings for the Z-factor and velocity renormalization, it is deduced that, the larger electron lifetimes result in the reduced scattering rates. Hence, the intra-band scattering is the more important transition process for these two modes. For the first effect, as it is stated from the Z-factor results, the lifetimes of electrons are affected by the localized polaronic states consequently, in these optical modes inter-band scattering processes are relevant.
Accordingly, to have low thermal conductivity $\tau_{ph-ph}$ (the phonon-phonon scattering life time which is an anharmonic effect) should be small and $\tau_{e-e}$ (the bare e-e scattering life time) should be large. Referring to Fig.\ref{lifetime2}, it is depicted that a small $\tau_{ph-ph}$ is relevant with small $\tau_{e-ph-e}$ or equivalently, large values of $Im\sum_{e-ph-e}$ which elucidated that the optical modes have the dominant part in low thermal conductivity of cubic zirconia.

\subsection{CONCLUSIONS}

In summary, we applied the $G_0W_0 @DFPT$ method within the frozen phonon model to analyze the phonon-assisted electron dynamics of cubic $ZrO_{2}$. We have traced phonons by observing the indirect electron-electron interaction to refrain from the time consuming calculations associate with the direct approach. Despite the strong phonon-phonon interactions, from the obtained indirect electron-phonon coupling matrix along with the carrier velocity and energy renormalization, we observed a substantial polaronic character for all the activated lattice vibrations through the optical phonons except for the second TO and the first LO modes, making these two dynamical configurations appropriate for practical purposes. Furthermore, we observed a small to large polaron transition in the first LO mode. Our many-body GW calculations through the frozen phonon model, pave the way for the simple estimation of the order of electron-phonon and phonon-phonon interactions.

\subsection{ACKNOWLEDGMENTS}

This work was supported by the Vice Chancellor for Research Affairs of Yazd University. The authors appreciate Nano Structured Coatings Institute of Yazd Payame Noor university for providing the computational facilities.

\section{References}
\bibliography{zro2bib}

\end{document}